\documentclass[11pt,twoside]{article}
\usepackage{jltp_latex2e}
\usepackage{epsfig}
\usepackage{amsmath}
\title{Quantum-Monte-Carlo Calculations for Bosons in a Two-Dimensional
  Harmonic Trap}

	\author{Stefan Heinrichs$^{\dag}$\address{$^{\dag}$Department of 
	Physics, University of Konstanz, Germany} and William J. 
	Mullin\address{Department of Physics and Astronomy, Hasbrouck 
	Laboratory, University of Massachusetts, Amherst MA 01003-3720, 
	USA}}

\runninghead{S. Heinrichs and W. J. Mullin}{PIMC Calculations for Bosons in a
  2D Harmonic Trap}
       
\begin{document}
\newcommand{\eps}{\varepsilon}
\newcommand{\vphi}{\varphi}
\newcommand{\Tr}{\operatorname{Tr}}
\newcommand{\bessj}{\operatorname{J}}
\newcommand{\bessn}{\operatorname{N}}
\newcommand{\mkmath}[1]{\ifmmode{#1}\else{\mbox{$#1$}}\fi}
\newcommand{\mb}[1]{\mkmath{\mathbf{#1}}}
\newcommand{\imb}[1]{_{\mathbf{#1}}\mathnormal}
\newcommand{\bd}[1]{\textbf{#1}}
\newcommand{\emp}[1]{\textsl{#1}}

\newcommand{\gbox}[2]{\begin{center}
\framebox[#1cm]{\rule{0cm}{#2cm}}
\end{center}}
\begin{abstract}
Path-Integral-Monte-Carlo simulation has been used to calculate the
properties of a two-dimensional (2D) interacting Bose system.  The
bosons interact with hard-core potentials and are confined to a
harmonic trap.  Results for the density profiles, the condensate
fraction, and the superfluid density are presented.  By comparing with
the ideal gas we easily observe the effects of finite size and the
depletion of the condensate because of interactions.  The system is
known to have no phase transition to a Bose-Einstein condensation in
2D, but the finite system shows that a significant fraction of the
particles are in the lowest state at low temperatures.

PACS numbers:03.75.Fi,02.70.Lq,05.30.Jp,05.70.Fh,32.80.Pj,67.40.Db
\end{abstract}
\maketitle
\section{INTRODUCTION}
Recent experiments on alkali atoms cooled by laser methods and 
evaporation in a magnetic trap have allowed the observation of 
Bose-Einstein condensation (BEC) in a harmonic potential formed by an 
external magnetic field.  By varying the trapping field so that it is 
very narrow in one dimension, it is possible to separate the 
single-particle states in the oscillator potential into well-defined 
bands.  By occupying only states in the lowest band one has an 
effective two-dimensional system.  In contrast to such a quasi-2D 
system, the system considered here is genuinely two dimensional.

In this paper we will investigate the behaviour of harmonic Bose systems in 2D
using the very powerful finite-temperature Path-Integral Monte Carlo (PIMC)
simulation technique\cite{Ceperley}. The PIMC technique is in principle
capable of describing systems of arbitrary interaction strength and density
and allows one to study the static properties of the condensed gases. The only
fundamental uncertainty arises from the choice of the interaction potential.
Here a hard-core potential appropriate for the s-wave scattering length of
$^{87}$Rb was chosen. In this case the hard-core parameter is $a_0=0.0043$ in
the dimensionless units of Ref. 2. The analogous three dimensional
case has been studied previously\cite{Krauth}.

The Hamiltonian for the system under consideration is
\begin{equation}
  \label{eq:syst-hamilt}
  H = \sum_{i=1}^{N} \frac{p_i^2}{2 m} + \sum_{i,j} V(r_i - r_j) + \frac{m}{2}
  \sum_{i=1}^{N} (\omega_x^2 r_{i,x} + \omega_y^2 r_{i,y}).
\end{equation}

The density matrix for $N$ Bose particles at inverse temperature
$\beta=1/k_bT$ can be written as a convolution with $M$ intermediate density
matrices or "time slices" at inverse temperature $\tau = \beta/M$. Then the
probability density for finding a many-particle configuration $\mb{R}=(r_1,
\cdots, r_N)$ is
\begin{equation}
\begin{split}
\rho (\mathbf{R},\mathbf{R};\beta )
&= \frac{1}{N!} \sum_{\mathcal{P}} 
\int \cdots \int \rho (\mathbf{R},\mathbf{R}_{1};\tau ) \times \\
& \quad \quad \times \rho (\mathbf{\ R}_{1},\mathbf{R}_{2};\tau )\cdots 
\rho (\mathbf{R}_{M-1},\mathbf{R}^{\mathcal{P}};\tau ) 
d\mathbf{R}_{1}d\mathbf{R}_{2}\cdots d\mathbf{R}_{M-1}  \nonumber
\end{split}
\end{equation}
where $\mathbf{R}^{\mathcal{P}}$ denotes a vector with permuted particle
labels. The Metropolis algorithm is used to sample from this distribution.
With larger M or correspondingly larger temperatures for each time slice, the
intermediate density matrices approach the classical limit which leads to the
{\em primitive approximation}. In order to make the computation for many
particles feasible, $M$ can be reduced by several orders of magnitude by
calculating the density matrix $\rho_2$ for the interaction involving just two
particles and approximating each time slice by\cite{Ceperley}
\begin{equation}
  \label{eq:rho-prod}
 \rho(\mb{R},\mb{R}';\tau) = \prod_{i=1}^N \rho_{1}(r_{i}, r_{i}';\tau) 
 \prod_{i<j} \frac{\rho_{2}(r_{i},r_{j};r_{i}',r_{j}';\tau)}
{\rho_{1}(r_{i},r_{i}';\tau) \rho_{1}(r_{j},r_{j}';\tau)}.
\end{equation}
As long as only two particles interact, this is equivalent to using the
primitive approximation.

To calculate $\rho_2$, we note that in the harmonic potential the Hamiltonian
for two particles decouples into a center of mass and a relative motion term
with the latter given by
\begin{equation}
  \label{eq:rel-hamilt}
  H_{rel} = \underbrace{\frac{p^2}{2 \mu} + V(r)}_{H_{HC}} + \frac{\mu}{2} 
                   (\omega_x^2 r_x^2 + \omega_y^2 r_y^2)
\end{equation}
where $\mu=m/2$ is the reduced mass. With the Trotter breakup, the quotient
in (\ref{eq:rho-prod}), which plays the role of a correction term, can be
transformed to
\begin{equation}
  \label{eq:corr}
   \rho_{HC}(r,r') \frac{X(r) X(r')}{\rho_{1,\mu}(r,r')}  
\end{equation}
with $X(r) = \exp(-\tau \mu (\omega_x^2 r_x^2 + \omega_y^2 r_y^2) / 4) $ and
the relative coordinates $r=r_i - r_j$ and $r'=r'_i-r'_j$. $\rho_{HC}(r,r')$
is the density matrix for the Hamiltonian $H_{HC}$ (Eq.~(\ref{eq:rel-hamilt}))
describing the interaction between two interacting particles without confining
potential and $\rho_{1,\mu}(r,r')$ is the density matrix for a single particle
of reduced mass $\mu$ in the harmonic potential\cite{Feynman}.

Calculating $\rho_{HC}(r,r')$ for a hard-core potential using an eigenfunction
expansion yields
\begin{equation}
 \rho_{HC}(r,r';\tau) = \sum_{l=-\infty}^{\infty} \frac{1}{2 \pi}e^{il\vphi}
\int_0^{\infty}dk R_{kl}(r) R_{kl}(r') e^{-\tau \frac{\hbar^2 k^2}{2\mu}}
  \label{rho-2p}
\end{equation}
with the radial wave functions 
$$
R_{kl}(r) = \sqrt{k}(\cos\delta_{kl} \bessj_l(kr) - \sin \delta_{kl}
\bessn_l(kr)) \quad \text{and} \quad \tan \delta_{kl} =
\frac{\bessj_l(ka_0)}{\bessn_l(ka_0)}$$
where $\bessj_l$ and $\bessn_l$ are
the Bessel and Neumann functions and $\vphi$ is the angle between $r$ and
$r'$.

Since the numerical evaluation of this function is quite time 
consuming, values distributed over a mesh with parameters $|r|, |r'|$ 
and $\vphi$ are computed for each value of $\tau$.  A simple linear 
interpolation method proved to be sufficient to reach the required 
degree of accuracy for evaluation in the simulation.  The full 
correction factor (Eq.~(\ref{eq:corr})) with dependence on the 
confining potential can then be calculated very efficiently during the 
Monte-Carlo simulation.

To further speed up the calculation, a boxing algorithm\cite{Krauth} is used that
divides the space of the system into boxes with a size of at least the ``healing
length'' of the pair interaction. When we compute the effects of the interaction
of a given particle, it is  necessary to consider only a small number of
interactions with particles in boxes neighbouring that of the particle in
question. Furthermore, attempting only permutations with particles from the
same box, increases the acceptance probability for these moves and results in
a more effective sampling of permutation space.\cite{Krauth}

The value of $\tau$ has to be chosen carefully for each particle
density to accommodate both the Trotter breakup and the approximation of pair
interactions (Eq. (\ref{eq:rho-prod})). For the density used, tests with values
for $\tau$ ranging over orders of magnitude have been performed, showing that
a further decrease below $\tau = 0.01$ yield the same results within
statistical errors. 

\section{DENSITY PROFILES AND CONDENSATE FRACTION}
In 2D, with interactions, there is no phase 
transition to a Bose condensed state\cite{Mullin} even in a trap, 
nevertheless there should be a macroscopic fraction of particles in 
the ground state at finite temperature when the particle number is 
finite.

We tentatively assume that, in 2D, the condensation fraction can be 
determined, as in 3D\cite{Krauth}, by observing a macroscopic number 
of particles with long exchange cycles in the lowest state of their 
subsystem.  For each temperature a characteristic length of 
permutation cycles $l_0$ can be chosen so that the density profiles 
for particles on cycles longer than $l_0$ are essentially the same.  
Particles on shorter cycles will have a broader density profile.  
Particles on cycles longer than $l_0$ are identified with the 
condensate (Fig.~1).  What is plotted is the average distribution in 
one coordinate after integrating over the other coordinate.  We have 
fit the lowest temperature curve in Fig.~1 with the infinite-N 
solution of the zero-temperature GP equation\cite{BP}.  The 
interaction coefficient has been determined by the fit since, in 2D, 
there is no straightforward connection between a hard-core interaction 
and a contact pseudopotential.  We observe no obvious bimodality, due 
to separate distributions of condensed and non-condensed particles in 
the overall density in Fig.~1; if a kink did exist in the density it 
would likely be washed out by the integration over one coordinate.

In two dimensions interactions seem to lead to a comparatively 
stronger depletion of the condensate (Fig.~2) than in three 
dimensions (Cf., Ref.~3).  This is expected, since there is no 
condensate at finite temperature in the thermodynamic limit for the 
interacting system and therefore the critical temperature has to 
approach zero (for large numbers of particles) when interactions are 
turned on.\cite{Mullin}


\section{SUPERFLUID FRACTION}
Recently one of the authors \cite{Mullin2} showed that the 
Hartree-Fock-Bogoliubov equations indicated that there is a phase 
transition in 2D, although it cannot be to the BEC state.  Perhaps 
this transition is of the Kosterlitz-Thouless (KT) type 
\cite{Shevchenko} and involves superfluidity.  The superfluid fraction 
can be related to the mean square surface area enclosed by Feynman 
paths.  Sindzingre \emph{et.~al.}\ have shown that\cite{Szin}
\begin{equation}
\rho_s/\rho = \frac{4m^2 \langle A^2 \rangle}{\beta h^2 I_c}
\end{equation}
with $A$ the area swept out by the paths and $I_c$ the classical moment of
inertia. The average is taken over configurations in the simulation. 

The resulting superfluid densities are very small (Fig.~2). A
smaller $\rho_s$ than in the translationally invariant system is expected
because the formation of the superfluid begins in the middle of the potential
well where the contribution to the moment of inertia is small. Paths with different
orientations will contribute with different signs to the area making
cancellation possible.

The simulation results for the 2D system without confining
potential\cite{2d-wp} are in agreement with the KT
theory for the superfluid transition. If this description is also appropriate
with confining potential, the vortex picture of the KT-transition suggests an
additional mechanism leading to a decrease of $\rho_s$\cite{Shevchenko} :
Superfluidity is destroyed by dissipation through vortices which are not
paired. At low temperatures vortices form pairs which unbind at higher
temperatures. In the non-uniform system the two vortices forming a pair will
in general experience a slightly different potential leading to imperfect
pairing and thereby a lowering of the superfluid fraction.

\begin{flushleft}
\epsfxsize = 5 in
\epsfbox{density.epsf}
\end{flushleft}
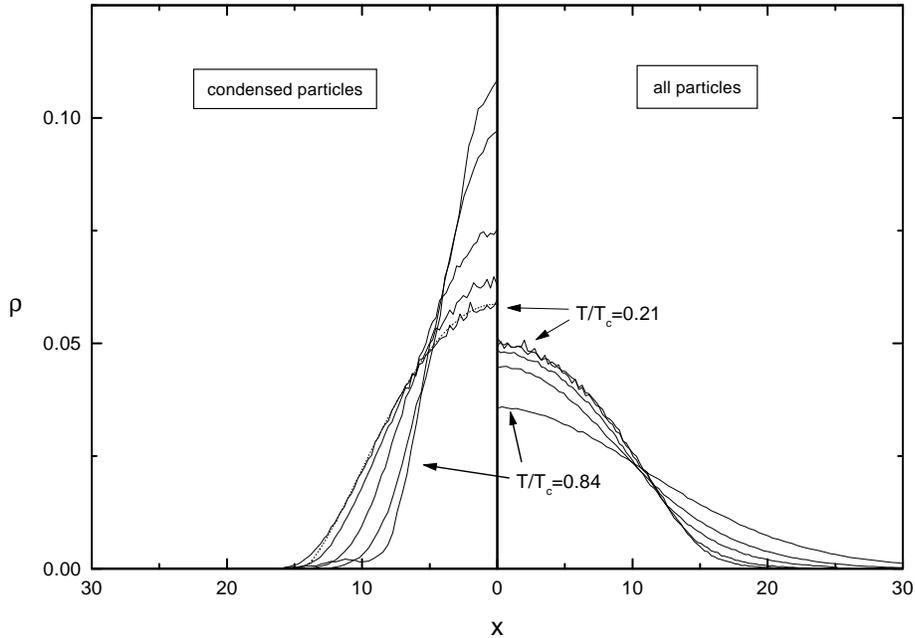
Fig.~1. Density profiles for 1000 particles at temperatures
$T/T_c=0.84, 0.56, 0.42, 0.28, 0.21$ for $\omega = 0.15$.  The 
condensate parts are displayed on the left and the profiles for all 
particles on the right.  All profiles are normalised to unity.  Temperatures are 
given with reference to the critical temperature $T_c$ of the ideal 
system in the thermodynamic limit. The smooth dotted curve through the 0.21 
data is the infinite-N solution of the GP equation with 
interaction strength determined by the fit.
 
\begin{flushleft}
\epsfxsize = 4.75 in
\epsfbox{cf-sf.epsf}
\end{flushleft}
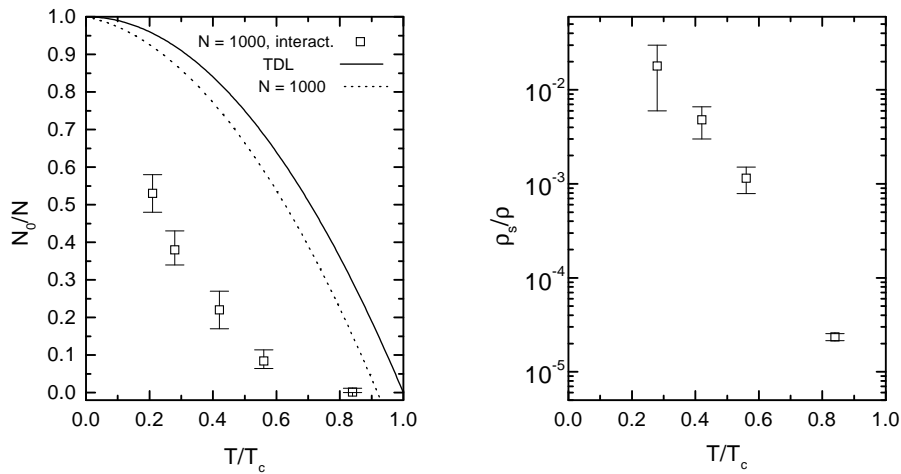
Fig.~2. Left figure: Condensate fraction for the interacting system 
with N=1000 and $\omega = 0.15$ and comparison with theoretical 
prediction for the ideal gas in the thermodynamic limit (TDL) and with 
finite size corrections for 1000 particles. Right figure: 
Dependence of the superfluid fraction on temperature for 1000 
particles.

\section*{ACKNOWLEDGEMENTS}
We thank Werner Krauth for providing his PIMC program for the three
dimensional case\cite{Krauth}.


\begin{thebibliography}{9}
  
\bibitem{Ceperley} D. Ceperley, {\it Rev.\ Mod.\ Phys.} {\bf 67}, 279 (1995);
\\
  E. L. Pollock, and D. M. Ceperley, {\it Phys.\ Rev.\ B} {\bf 30}, 2555 (1984).

\bibitem{units} F. Dalfovo, S. Stringari, {\it Phys. Rev. A}, {\bf 53}, 2477 (1996).
  
\bibitem{Krauth} W. Krauth, {\it Phys. Rev. Lett.} {\bf 77}, 3695 (1996).
  
\bibitem{Feynman} R. P. Feynman A. R. and Hibbs, {\it Quantum Mechanics and
    Path Integrals } (McGraw-Hill Inc., New York 1965).
\\
  R. P. Feynman, {\it Statistical Mechanics}, (Benjamin, New York, 1972) .

\bibitem{Mullin} W. J. Mullin, {\it J. Low Temp. Phys}, {\bf 106}, 615 (1997).

\bibitem{BP} G. Baym and C. Pethick, {\it Phys. Rev Lett.}, {\bf 76}, 6 (1996).

\bibitem{Mullin2} W. J. Mullin, {\it J. Low Temp. Phys}, {\bf 110}, 167 (1998).

\bibitem{Shevchenko} S. I. Shevchenko, {\it Sov. Phys. JETP}, {\bf 73}, 1009
  (1991).
  \\
  S. I. Shevchenko, {\it Sov. J. Low. Temp. Phys.}, {\bf 18}, 223 (1992).

\bibitem{Szin} P. Sindzingre, M. Klein, and D. M. Ceperley, {\it Phys.  Rev.
    Lett.} {\bf63}, 1601 (1989).

\bibitem{2d-wp} D. M. Ceperley and E. L. Pollock, {\it Phys. Rev. B} {\bf 39},
  2084 (1989).
  

\end{thebibliography}
\end{document}